\documentstyle[11pt]{article}
\date{}
\begin{document}
\sloppy
\title{Is the negative vacuum energy around galaxies
a possible candidate for dark matter?}
\author{V. Majern\'{\i}k \\
Institute of Mathematics, Slovak
Academy of Sciences, \\ SK-81473 Bratislava, \v Stef\'anikova  47,
 Slovak Republic\\ and\\
Department of Theoretical Physics\\ Palack\'y
University\\T\v r. 17. listopadu 50\\772 07 Olomouc, Czech Republic}
\maketitle
\begin{abstract}
\noindent
In this short communication we
point out to the possibility that the clouds of the
negative vacuum energy around galaxies may have the same effect as dark matter.
These clouds amplify the force acting on the  baryonic matter around
galaxies changing their dynamical and
kinematical properties. Inserting the density of the baryonic matter
into field equations of the cosmic quaternionic field for the present
time yields the negative cosmological constant
which is approximately equal to the density of the baryonic matter of
galaxies. Due to the effect of the negative vacuum energy on moving
cosmical objects the effective total masses of galaxies assume several
times larger values as their baryonic masses.
 \footnote{E-mail:majere@prfnw.upol.cz}
\end{abstract}

\section{Introduction}

The recent astronomical observations
\cite{PER} \cite{pe} \cite{BO} \cite{T} give increasing
support for the accelerating and flat universe which
consists of a mixture of a small part of the
baryonic matter about one third non-relativistic dark matter (DM) and two
thirds of a smooth component, called dark energy (DE).
In this communication, we put forward the thesis that DM is
a type  of the negative
vacuum energy concentrated around galaxies which can be modelled by
{\it negative} cosmological constant $\Lambda$ whose value can be
determined by means of the field equations of the cosmic quaternionic
fields.

In the literature,
DE is theoretically modelled by many ways, e.g. as
(i) a very small cosmological constant (e.g.\cite{4}) (ii) quintessence
(e.g.\cite{5})
(iii) Chaplygin gas (e.g.\cite{6}) (iv) tachyon field (e.g.\cite{7}
\cite{PA} \cite{PAD}) (v)
interacting quintessence (e.g.\cite{8}),
 quaternionic field (e.g.\cite{IK}),
etc.
It is unknown which of the said models will finally
emerges as the successful
one.

Another fundamental problem being faced to cosmology
is that of the nature of DM, which is supposed to exist because of
dynamical astronomical measurement but we have not yet detected it.
Astronomers found that one third of the universe's mass is made up of unknown
matter that is invisible to telescopes but have gravitational
effects on the baryonic matter \cite{D}.
Lacking evidence of direct detection, the presence, nature, and quantity
of DM must be inferred from the kinetic and distribution
properties of baryons.
In the literature, many sophisticated candidates of
DM have been proposed
which can be divided up into two categories. Some authors proposed
that DM consists of a kind of matter substance, e.g.
of the massive particles (WIMPs)
which stems from extensions to the standard model of particle physics,
such as supersymmetry and extra-dimensional theory. Other assume
that DM represents
the relics of
primordial black holes (see, e.g.\cite{F}).
On the other side, several authors try explain the kinematic effects
assigned to DM by modifying of Newton's law of
gravitation (see,e.g. \cite{P}).

In what follows, we will show that clouds of negative vacuum energy around
galaxies enhance the gravitation force acting on cosmological objects
which has a similar effect on their rotation curves as if one would
add to their baryonic mass densities an additional masses.
These 'additional masses' could represent the corresponding
DM seemly concentrated around galaxies.

\section{The field equations for the cosmic quaternionic field}

In a recent article \cite{MK},
$\Lambda$ has been interpreted as the {\it field energy} of the
cosmic quaternionic field (called $\Phi$-field, for short).
The field  equations
for the $\Phi$-field can be written in the following form (c=1)
$$\partial_i F_{ij}=J_j, \quad \eqno(1)$$
where $J$ is the 4-current of ordinary matter with  the components
$J_j=k\rho v_j$
(j=1,2,3) being the components of space velocity and  $\rho$ is the
matter density;
$J_0=k(\epsilon_{self}+\rho)$ where $\epsilon_{self}$ is the energy
density of the $\Phi$-field.
$F_{ij}$ has only diagonal
components $F_{ii}=\Phi,\quad i=1,2,3,\quad F_{ii}=-\Phi, \quad
i=0,$ and $F_{ij}=0,\quad i\neq j$.
The field
variable $\Phi$ has the dimension of field
strength and its square has the dimension of energy density.
In the differential form the field
equation (1) becomes (c=1)
$$\nabla \Phi=k \vec J= k\rho \vec v,\quad\quad v_i, i=1,2,3\quad \eqno(2a)$$
and
$$-{\partial \Phi \over c \partial
t}=k_0(\epsilon_{self}+\rho),
\quad\quad
i=0 \quad\eqno(2b)$$
where $\epsilon_{self}$ is the energy density of the $\Phi$-field given
as \cite{X}
$$\epsilon_{self}=\frac{\Phi^2}{8 \pi}.$$
If one writes Newton's law of gravitation
in the symmetrical form
$$F=-\frac{(q_g)^2}{r^2}=-\frac{(\sqrt{G}m)^2}{r^2},$$
then the quantity $q_g=\sqrt{Gm}$ can be understood as the
'gravitational charge', assigned to the mass
$m$ \cite{BR} \cite{JM}. Next, we will use gravitational charges when
formulating the field equations of the cosmic quaternionic field.
Since the r-h-side of Eqs.(2a) and (2b) oughts to have the
dimension of gravitational charge we set for the coupling constants $k$ and
$k_0$ as
$$k=\sqrt{G}\qquad k_o=8\pi \sqrt{G}.$$
Inserting $k$ and $k_0$ into Eqs. (2a) and (2b) we have
$$\nabla \Phi=\sqrt{G} \vec J= \sqrt G \rho \vec v,\quad\quad v_i, i=1,2,3\quad
\eqno(2a^{'})$$
and
$$-{\partial \Phi \over c \partial
t}=8 \pi \sqrt{G}(\epsilon_{self}+\rho),
\quad\quad
i=0 . \quad\eqno(2b^{'})$$
These equations are first-order differential equations whose
solution can be found given the source terms.
According of the boundary conditions we distinguish three types of
solutions of Eqs. (2a') and (2b'):\\
(i) For $\rho =0$, the only source of the $\Phi$-field
is its {\it own} energy density, i.e. $\epsilon_{self}=\Phi^2/8\pi$.
Assuming the spacial homogeneity of the $\Phi$-field and the absence of
any ordinary matter then the field variable $\Phi$
will be dependent only on time $t$.
Therefore, Eqs.(2a) and (2b) become
$$\nabla\Phi= 0\quad \eqno (3a)$$
and
$$-{d \Phi \over d t}= \sqrt {G}\Phi^2.\quad\eqno(3b)$$
The solution of (3b) has a
simple form
$$\Phi(t)= {1 \over \sqrt {G} (t+t_0)},$$
where $t_0$ is the integration constant given by the boundary
condition.\\
(ii) For $|\vec v|\ll c$ and $\rho \neq 0$ we have
$$\nabla\Phi\approx 0\quad \eqno (4a)$$
and
$$-{d \Phi \over d t}= 8\pi \sqrt
{G}\left(\frac{\Phi^2}{8\pi}+\rho\right ).\quad\eqno(4b)$$
For the present time $t\approx 10^{10} yr$, the
derivation of $d\Phi/dt$  become approximately equal zero
and Eq. (4b) turns out to be
$$ 8\pi \sqrt
{G}\left(\frac{\Phi^2}{8\pi}+\rho\right )= 8\pi \sqrt
{G}\left(\Lambda+\rho\right ) \approx 0.\quad\eqno(5)$$
The {\it field energy density} of the cosmic
quaternionic field $\Phi/(8\pi)$ for ($t_0=0$) is
$$\Lambda=\frac{\Phi^2}{8\pi}=
\frac{\lambda}{8\pi }\left [\frac{1}{\sqrt{G}t}
\frac{1}{\sqrt{G}t}\right ] .$$
From Eq.(5) it follows
$$\Lambda\approx -\rho.\quad\eqno (7)$$
For a weak gravitation field of the spherically symmetric galaxy
the acceleration of a cosmic object is
given as \cite{H}
$$\ddot{R}=-\frac{G\int_{0}^R{\rho(r) r^2 4\pi
dr}}{R^2}+\frac{R\lambda}{3}.\quad\eqno(8)$$
Keeping in mind that $\lambda =8\pi G \Lambda$ and $\Lambda\approx
-\rho$, respectively,
we have finally
$$A=\ddot R=A_1+A_2=-\frac{G\int_{0}^R{\rho(r) r^2 4\pi dr}}{R^2}-\frac{8 \pi G
R\rho}{3}.\quad\eqno(9)$$
We see that the first and second
term in Eq. 8 represents the gravitation acceleration and the acceleration
due to negative cosmological constant, respectively.
Equaling the centripetal and the acceleration A, the velocity of the
rotational curves as a function of $\rho$ and $R$ becomes
$$v_R=\sqrt{\frac{G\int_{0}^R{\rho(r) 4\pi r^2 dr}}{R}+\frac{8 \pi
 G \rho R^2}{3}}. \quad\quad\eqno(10)$$
 
For example, consider a galaxy with the mass distribution
$\rho =const=\rho_0$.
The velocity of the rotational curve $v_R$ is\\
$$v_R=\sqrt{\frac{G\int_{0}^R{\rho_0(r) r^2 4\pi dr}}{R}+\frac{8 \pi G
\rho_0 R^2}{3}}=
R\sqrt{G\left(\frac{4 \pi \rho_0}{3} \right )\left(1+2)\right)}$$
while
without the presence of the negative vacuum energy this velocity is
$$v_1'=R\sqrt{G\left (\frac{4 \pi \rho_0}{3} \right )} . $$
We see the presence of the negative vacuum energy is the same as if the
density of galaxy were $3\rho_0$, i.e. as if the total mass of galaxy
would increases.
The clouds of negative vacuum energy
around galaxies behaves as DM. Yet, this does not excludes that also other
components of DM exists.

\section{Consequences}

It is generally believed that
most of energy and matter in our universe is of unknown nature to us,
therefore  to explain  the nature of DM is one of the major fundamental
challenges of present astrophysics. In this communication
we have attempted to show that the
cloud of the negative vacuum energy around galaxy whose
 density is approximately equal to $\rho$ has the same effect
as DM occurring in them.\\

From what has been said so far the following points are worth of
mentioning\\
(i) The clouds of negative vacuum energy around
galaxies has the same effect as presence of additional mass in them.\\
(ii) The density of the negative  vacuum energy
is approximately equal to
the mass density of galaxies.\\
(iii) The whole kinematics of star motion in galaxies
is given by their baryonic mass distribution.\\
(iv) It seems probable that DM does not exists {\it per se}, but
DM could just be some type of vacuum energy concentrated around
galaxies (see also \cite{B}).\\
(v) The density of the negative vacuum energy can be determined by means
of field equations of the cosmic quaternionic field.

The aim of this short communication was only to outline the basic idea of a
possible new interpretation of dark matter. Many important issues remained
open and may be eventually solved by the future detailed theory and
its application to a realistic model of galaxies.

\end{document}